\documentclass[traditabstract]{aa}
\usepackage{amsmath}
\usepackage{txfonts}
\usepackage{graphicx}
\usepackage{natbib}
\bibpunct{(}{)}{;}{a}{}{,}

\begin{document}

\title{The M dwarf planet search programme at the ESO VLT + UVES
\thanks{Based on observations collected at the European Southern Observatory, Paranal Chile, ESO programmes 65.L-0428, 66.C-0446, 267.C-5700, 68.C-0415, 69.C-0722, 70.C-0044, 71.C-0498, 072.C-0495, 173.C-0606, 078.C-0829.}
\thanks{Radial velocity data are available in electronic form at the CDS via anonymous ftp to cdsarc.u-strasbg.fr (130.79.128.5) or via http://cdsweb.u-strasbg.fr/cgi-bin/qcat?J/A+A/}
}

\subtitle{A search for terrestrial planets in the habitable zone of M dwarfs}

\author{M. Zechmeister\inst{1}\and M. K\"{u}rster\inst{1}\and M. Endl\inst{2}}

\institute{Max-Planck-Institut f\"{u}r Astronomie, K\"{o}nigstuhl 17, 69117 Heidelberg, Germany\\ \email{zechmeister@mpia.de}\and
McDonald Observatory, University of Texas, Austin, TX78712, USA}

\date{Received / Accepted}

\abstract{We present radial velocity (RV) measurements of our sample of 40 M dwarfs from our planet search programme with VLT+UVES begun in 2000. Although with our RV precision down to 2--2.5\,m/s and timebase line of up to 7 years, we are capable of finding planets of a few Earth masses in the close-in habitable zones of M dwarfs, there is no detection of a planetary companion. To demonstrate this we present mass detection limits allowing us to exclude Jupiter-mass planets up to 1 AU for most of our sample stars. We identified 6 M dwarfs that host a brown dwarf or low-mass stellar companion. With the exception of these, all other sample stars show low RV variability with an rms$\,<20\,$m/s. Some high proper motion stars exhibit a linear RV trend consistent with their secular acceleration. Furthermore, we examine our data sets for a possible correlation between RVs and stellar activity as seen in variations of the H$\alpha$ line strength. For Barnard's star we found a significant anticorrelation, but most of the sample stars do not show such a correlation.}


\authorrunning{Zechmeister et al.}

\maketitle

\section{Introduction}

High-precision differential radial velocity (RV) measurements of stellar reflex motions induced by an orbiting companion have so far been the most successful method to discover extrasolar planets and to characterise their orbital properties. Originally, RV planet search programmes have largely concentrated on main sequence stars of spectral types late-F through K. That only a comparatively small number of M~dwarfs were included comes from their faintness, which requires large telescopes to perform high-precision RV measurements of a few m/s. For an understanding of the formation and abundance of extrasolar planets it is important to determine the presence and orbital characteristics of planets around stars of as many different types as possible, and especially around this most abundant type of star.

Even if M~dwarfs are faint and require large telescopes, however, they have two advantageous characteristics when searching for terrestrial exoplanets in the habitable zones (HZ) with radial velocity (RV) methods. Compared to solar-like stars, (i) they have a lower mass ($M=0.1-0.6M_{\odot}$), and (ii) the habitable zone is close-in around this cooler and less luminous type of star ($L=0.0008-0.06L_{\odot}$, \citealt{Guinan09}). By habitable zone we understand the region that allows liquid water on the planet surface as described in \citet{Kasting93}.

For M~dwarfs, the HZ is typically 0.03 -- 0.4 AU. The RV amplitude induced due to a planet by the Doppler effect is
\begin{equation}
K=\sqrt{\frac{G}{1-e^{2}}}\frac{m\sin i}{\sqrt{(M+m)a}}=28.4\mathrm{m/s}\cdot\frac{m\sin i}{M_{\mathrm{Jup}}}\left(\frac{M_{\odot}}{M}\frac{\mathrm{AU}}{a}\right)^{1/2}.\label{eq:Amplitude}
\end{equation}
It increases for closer distances $a$ (shorter periods) and lower stellar mass $M$. Thus the RV amplitude induced by a planet in the HZ of an M~dwarf is higher than that of a solar-like star. M~dwarfs are also ideal targets for astrometric follow-up due to their lower mass, as well as for transit observations. In combination with the RV method, astrometry allows the resolution of the $\sin i$-ambiguity and true masses to be obtained.

\begin{table}
\caption{\label{Tab:Mdwarfplanets}M~dwarfs with known planets and their masses discovered with the RV method.}
\begin{tabular}{lccccl}
\hline
\hline Star & b & c & d & e & References\\
\hline
GJ~876 & $2.53M_{\mathrm{Jup}}$ & $0.79M_{\mathrm{Jup}}$ & $7.53M_{\oplus}$ & & {[}1{]}{[}2{]}{[}3{]}{[}4{]}\\
GJ~581 & $15.6M_{\oplus}$       & $5.1M_{\oplus}$        & $8.2M_{\oplus}$  & $1.9M_{\oplus}$ & {[}5{]}{[}6{]}{[}7{]}\\
GJ~436 & $21M_{\oplus}$         & & & & {[}8{]}\\
GJ~317 & $0.71M_{\mathrm{Jup}}$ & & & & {[}9{]}\\
GJ~674 & $11.1M_{\oplus}$       & & & & {[}10{]}\\
GJ~849 & $0.82M_{\mathrm{Jup}}$ & & & & {[}11{]}\\
GJ~176 & $8.4M_{\oplus}$        & & & & {[}12{]}{[}13{]}\\
GJ~832 & $0.64M_{\mathrm{Jup}}$ & & & & {[}14{]}\\
\hline
\end{tabular}
Note: The masses are minimum masses with the exception for GJ~436 and GJ~876. {[}1{]} \citealt{Delfosse98}, {[}2,3{]} \citealt{Marcy98,Marcy01}, {[}4{]} \citealt{Rivera05}, {[}5{]} \citealt{Bonfils05}, {[}6{]} \citealt{Udry07}, {[}7{]} \citealt{Mayor09}, {[}8{]} \citealt{Butler04}, {[}9{]} \citealt{Johnson07}, {[}10{]} \citealt{Bonfils07}, {[}11{]} \citealt{Butler06}, {[}12{]} \citealt{Forveille09}, {[}13{]} \citealt{Butler09}, {[}14{]} \citealt{Bailey09}.
\end{table}

At present there are only a few M~dwarfs known to have planets, as summarized in Table~\ref{Tab:Mdwarfplanets}. The M4V star GJ~876 has two Jovian planets orbiting in a 2:1 resonance (\citealt{Marcy01}; see also \citealt{Benedict02} for an astrometric determination of the mass of the outer planet), and a third planet in this system has been found by \citet{Rivera05}. The M3V star GJ~581 is another multiple system with three Neptune-type planets. The latest detection of a fourth planet in this system by \citet{Mayor09} is a new highlight in planet search with RVs. With only $1.9M_{\oplus}$, GJ~581~e is the lowest-mass planet found so far with the RV method. The same work also resulted in a revised period for the outer planet GJ~581~d, placing it inside the habitable zone. The planet around the M2.5V star GJ~436 first discovered with the RV method \citep{Butler04} also turned out to be a transiting one \citep{Gillon07}. So far it is the only known transiting planet around an M~dwarf. Most recently, \citet{Bailey09} has discovered the first long-period planet around an M~dwarf (GJ~832). A planet with a mass of $24.5M_{\oplus}$ ($P=10.24\,$d) around GJ~176 announced by \citet{Endl08} is rejected by \citet{Forveille09} and \citet{Butler09}. Instead, both groups find evidence of another planet with a shorter period (8.78~d) and a minimum mass of $8.4M_{\oplus}$ \citep{Forveille09} and $12M_{\oplus}$ \citet{Butler09}.

Several radial velocity (RV) surveys of M~dwarfs have resulted in few or no detections, indicating a lower frequency of planets compared to solar-like stars; e.g., \citet{Endl06} monitored 90 M~dwarfs (including the first data for 21 stars from our sample) without a planet detection. The sample studied by \citet{Cumming08} surveying 110 M~dwarfs contained only two planet hosting M~dwarfs (GJ~876 and GJ~436). HARPS guaranteed time project \#3 had 50 nights on 120 M~dwarfs within 11\,pc, and so far it has revealed only three planet-hosting M~dwarfs (GJ~876, GJ~581, and GJ~674).

M~dwarfs are an ideal testing ground for competing models of the formation of gas giants. While the classical core-accretion model has severe problems with forming Jupiter-mass planets in the less massive protoplanetary disks even around M~dwarfs (e.g. \citealt{Laughlin04}), the competing gravitational instability model can also efficiently form Jovian-type companions around M~dwarfs \citep{Boss06}. \citet{Ida05} even predict a higher frequency of icy giant planets with masses comparable to Neptune in short periodic orbits for M~dwarfs than for G type stars.

Recent results of microlensing surveys (e. g. OGLE-2005-BLG-390b, $5.5M_{\oplus}$, \citealt{Beaulieu06}; OGLE-2005-BLG-169b, \citealt{Gould06}; presumed M~dwarfs) may also indicate that low-mass planetary companions might be abundant around M~dwarfs.

\section{Targets and Observations}

Our sample consists of 40 M~dwarfs and one M~giant\footnote{GJ~4106 is listed in the Catalogue of Nearby Stars with a parallax of $84\pm17\,$mas (11.9\,pc). However, with the Hipparcos parallax GJ~4106 should be an M giant.}, which are listed in Table~\ref{Tab:Sample}. All M~dwarfs are brighter than $V\lesssim12.2\,\mathrm{mag}$ and nearby within a distance of 37\,pc (33 M~dwarfs even within 20\,pc). Their spectral types range from M0 to M5. The stellar masses~$M$ were derived from the mass-luminosity relation by \citet{Delfosse00} using the absolute brightness in K band. As an indicator of activity we selected the X-ray luminosity as detected in the ROSAT all-sky survey \cite[from][]{Huensch99}. Detected as X-ray sources are Barnard's star ($L_{x}=0.1$), GJ~1 (0.6), GJ~190 (0.9), GJ~229 (1.3), and the most active Proxima~Cen (1.7). The rest of the sample was not detected by ROSAT, implying that these stars are inactive.

The observations started in 2000, initially with 20 targets. Typically, three consecutive spectra per night were taken for each star with exposure times of 90 -- 900\,s depending on the object brightness. In April 2004 (JD=2453100) the sample was complemented by 21 additional stars, while the monitoring of HG~7-15 was ended. Since then the number of spectra per night was reduced to one, with the exception of Barnard's star, GJ~160.2, GJ~821, and Proxima~Cen.

Our data of the first 20 stars (+ GJ~510) taken before mid-2005 were already included in the study by \citet{Endl06}. Here we present the full data set for all 41 stars as observed until March 2007. A detailed study of the full Proxima~Cen data set can be found in \citet{Endl08b}. A study of the pre-2002.75 data set of Barnard's star is given by \citet{Kuerster03}.

\begin{table}
\caption{\label{Tab:Sample}Targets with their spectral type, visual magnitude~$V$, distance $d$ \citep{vanLeeuwen07}, and stellar mass $M$ derived from the K-band mass-luminosity relation by \citet{Delfosse00}.}
\centering
\begin{tabular}{llrrc}
\hline
\hline Star & Spec Type &
\multicolumn{1}{l}{$V$ {[}mag{]}}&
\multicolumn{1}{l}{$d$ {[}pc{]}}&
$M$ {[}$M_{\odot}${]}\\
\hline
Barnard~(GJ~699)  & M4Ve    &  9.54 &   1.82 & 0.16\\
GJ~1              & M1.5    &  8.57 &   4.34 & 0.45\\
GJ~27.1           & M0.5    & 11.42 &  23.99 & 0.53\\
GJ~118            & M2.5    & 10.70 &  11.65 & 0.36\\
GJ~160.2          & M0V     &  9.69 &  23.12 & 0.69\\
GJ~173            & M1.5    & 10.35 &  11.10 & 0.48\\
GJ~180            & M2      & 12.50 &  12.12 & 0.43\\
GJ~190            & M3.5    & 10.31 &   9.27 & 0.44\\
GJ~218            & M1.5    & 10.72 &  15.03 & 0.50\\
GJ~229            & M1/M2V  &  8.14 &   5.75 & 0.58\\
GJ~263            & M3.5    & 11.29 &  16.02 & 0.55\\
GJ~357            & M2.5V   & 10.85 &   9.02 & 0.37\\
GJ~377            & M3      & 11.44 &  16.29 & 0.52\\
GJ~422            & M3.5    & 11.66 &  12.67 & 0.35\\
GJ~433            & M1.5    &  9.79 &   8.88 & 0.48\\
GJ~477            & M1      & 11.08 &  18.99 & 0.54\\
GJ~510            & M1      & 11.05 &  16.74 & 0.49\\
GJ~620            & M0      & 10.25 &  16.44 & 0.61\\
GJ~637            & M0.5    & 11.36 &  15.88 & 0.41\\
GJ~682            & M3.5    & 10.96 &   5.08 & 0.27\\
GJ~739            & M2      & 11.14 &  14.09 & 0.45\\
GJ~817            & M1      & 11.48 &  19.17 & 0.43\\
GJ~821            & M1      & 10.87 &  12.17 & 0.44\\
GJ~842            & M0.5    &  9.74 &  11.99 & 0.58\\
GJ~855            & M0.5    & 10.74 &  19.15 & 0.60\\
GJ~891            & M2V     & 12.20 &  16.08 & 0.35\\
GJ~911            & M0V     & 10.88 &  24.26 & 0.63\\
GJ~1009           & M1.5    & 11.16 &  17.98 & 0.56\\
GJ~1046           & M2.5+v  & 11.62 &  14.07 & 0.40\\
GJ~1100           & M0      & 11.48 &  28.93 & 0.57\\
GJ~3020           & M2.5    & 11.54 &  22.78 & 0.62\\
GJ~3082           & M0      & 11.10 &  16.56 & 0.47\\
GJ~3098           & M1.5Vk: & 11.21 &  17.86 & 0.50\\
GJ~3671           & M0      & 11.20 &  17.74 & 0.50\\
GJ~3759           & M1V     & 10.95 &  16.97 & 0.49\\
GJ~3916           & M2.5V   & 11.25 &  15.10 & 0.49\\
GJ~3973           & M1.5Vk: & 10.94 &  18.23 & 0.54\\
GJ~4106           & M2      & 10.82 & 110.50 & 0.55\\
GJ~4293           & M0.5    & 10.90 &  25.06 & 0.57\\
HG~7-15           & M1V     & 10.85 &  37.31 & 0.78\\
Prox~Cen~(GJ~551) & M5.5Ve  & 11.05 &   1.30 & 0.12\\
\hline
\end{tabular}
\end{table}

\begin{figure*}[htp]
\centering
\includegraphics[height=0.96\textheight,width=1.\linewidth]{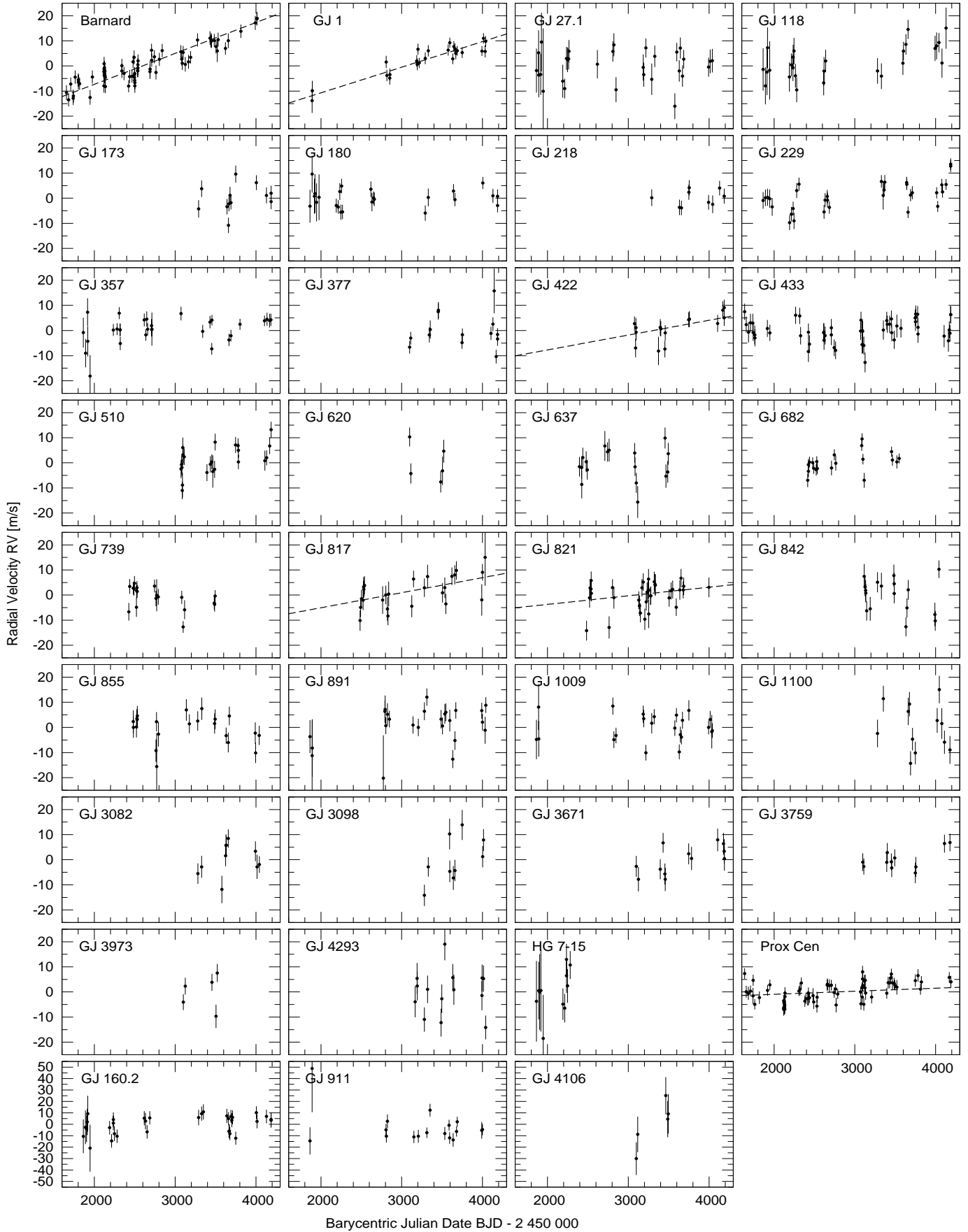}
\caption{\label{Fig:RV1}Radial velocities for 35 M~dwarfs. Dashed lines represent no fit; they show the predicted secular acceleration effect caused by the proper motion (only for Proxima~Cen and stars with $\dot{v}_{\mathrm{r}}>1\,$m/s/yr ).}
\end{figure*}

\begin{figure*}
\centering
\includegraphics[width=1\linewidth]{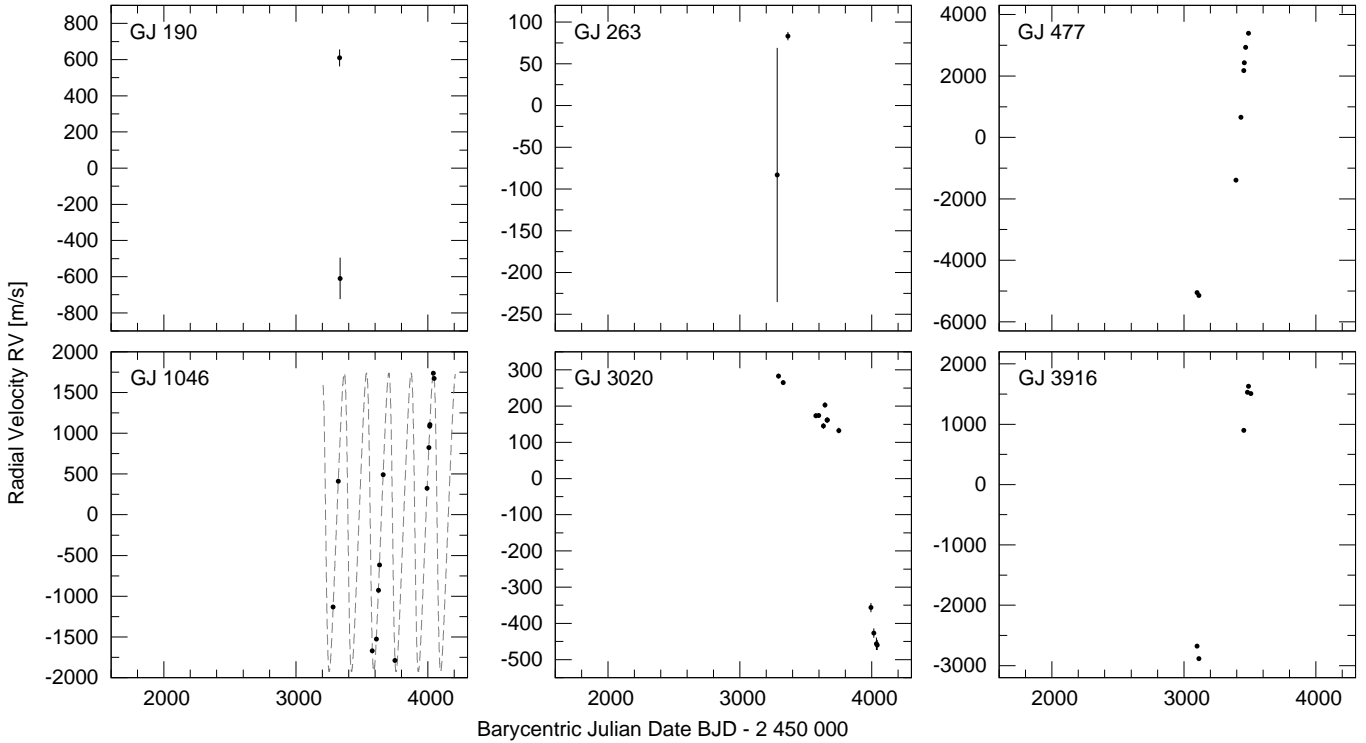}
\caption{\label{Fig:RV2}Radial velocities for 6 more M~dwarfs. The orbital solution for GJ~1046 is taken from \citet{Kuerster08}.}
\end{figure*}

The observations were carried out with the UVES spectrograph at the VLT-UT2, directly fed via image slicer \#3 that redistributes the light from a 1$\arcsec\times$1$\arcsec$ aperture along a 0.3$\arcsec$ wide slit. This resulted in a resolving power of $R=100\,000-120\,000$. The red arm of UVES was employed and a wavelength coverage 495 -- 704\,nm over 37 orders was selected. An iodine cell was used for precise wavelength calibration and modelling of the instrumental profile. Only the range of 500 -- 600\,nm was used to determine the RVs. This range is rich in iodine lines.

The data reduction included bias subtraction, flat-fielding, Echelle straylight subtraction, wavelength calibration, and barycentric correction \citep[see also][]{Endl08b}. The data modelling with the ``AUSTRAL'' code to obtain the RV is described in \citet{Endl00}. We achieve an RV precision of 2\,m/s for bright stars. In practice, photon noise limits the RV precision for faint M~dwarfs and correspondingly the errors are larger for those stars that were observed with lower S/N. Radial velocities and also their errors (see \citealt{Kuerster03} for discussion) were combined into nightly averages. They are shown in Figs.~\ref{Fig:RV1} and \ref{Fig:RV2} (the RV data are available as online material).

\section{Data analysis}

\subsection{Secular acceleration}

\begin{figure}
\includegraphics[width=1\linewidth]{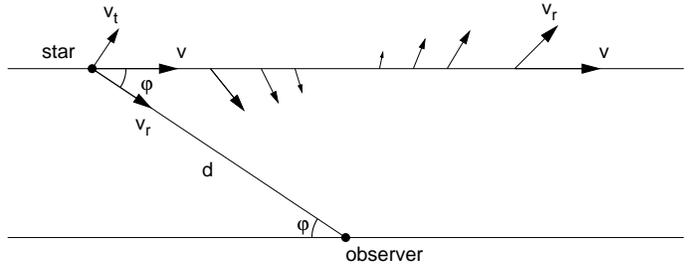}
\caption{\label{Fig:SecAcc}Change in the radial velocity $v_{r}$ due to constant motion $v$ (secular acceleration $\dot{v}_{r}$).}
\end{figure}

Even if a star moves undisturbed with a constant space velocity $v$, it can show a change in its radial velocity (RV). This secular or perspective acceleration was first measured by \citet{Kuerster03} for Barnard's star. In polar coordinates the radial and tangential velocity component, respectively, are (Fig.~\ref{Fig:SecAcc})
\[
v_{\mathrm{r}}=-v\cos\varphi\quad\text{and}\quad v_{\mathrm{t}}=v\sin\varphi.
\]
The differentiation of $v_{\mathrm{r}}$ with respect to time $t$ yields the secular acceleration ($v=\mathrm{const}.$)
\[
\dot{v}_{\mathrm{r}}=\frac{\mathrm{d}v_{\mathrm{r}}(t)}{\mathrm{d}t}=v\sin\varphi\cdot\dot{\varphi}=v_{\mathrm{t}}\dot{\varphi}.
\]
The tangential velocity $v_{\mathrm{t}}=d\cdot\mu$ depends on the distance $d$ of the star and its proper motion $\mu$, which can also be identified with the time derivative of the angular position $\dot{\varphi}=\mu$. Therefore the instantaneous secular acceleration is given by
\begin{equation}
\frac{\mathrm{d}v_{\mathrm{r}}(t)}{\mathrm{d}t}=\frac{v_{\mathrm{t}}^{2}}{d}=\mu^{2}d=22.98\frac{\mathrm{m/s}}{\mathrm{yr}}\frac{(\mu_{\alpha}^{2}+\mu_{\delta}^{2})\cdot\mathrm{yr}^{2}/\mathrm{arcsec}^{2}}{\pi/\mathrm{mas}}.\label{eq:SecuAcc}
\end{equation}
where $\mu_{\alpha}$ and $\mu_{\delta}$ are the proper motion in right ascension and declination, respectively. It only depends on the proper motion $\mu$ and parallax $\pi$, which are easily accessible from the Hipparcos catalogue. Note that the knowledge of $v_{\mathrm{r}}$ and $v$ is \emph{not} explicitly required for the prediction of the instantaneous secular acceleration.

One has to take this effect into account, especially for high proper motion stars, to avoid the misleading conclusion that this RV change is a disturbance by a companion. A feature of the secular acceleration is that this effect is only geometrical, not physical, and always positive ($\dot{v}_{\mathrm{r}}\ge0$). With the high astrometric precision of Hipparcos, this effect can be predicted with high accuracy for nearby stars.

\begin{table*}
\caption{\label{Tab:SecuAcc}Proper motion $\mu$, parallax $\pi$ (from Hipparcos, \citealt{vanLeeuwen07}) and secular acceleration (SA) $\dot{v}_{\mathrm{r}}(t)$ for the M~dwarfs sample, with the highest SA stars in the lower part of the table for comparison (not in our sample).}
\centering
\begin{tabular}{lrrrr}
\hline
\hline Star&
\multicolumn{1}{c}{$\mu_{\alpha}$}&
\multicolumn{1}{c}{$\mu_{\delta}$}&
\multicolumn{1}{c}{$\pi$}&
\multicolumn{1}{c}{$\dot{v}_{\mathrm{r}}(t)$}\\
&
\multicolumn{1}{c}{{[}mas/yr{]}}&
\multicolumn{1}{c}{{[}mas/yr{]}}&
\multicolumn{1}{c}{{[}mas{]}}&
\multicolumn{1}{c}{{[}m/s/yr{]}}\\
\hline
Barnard   &  -798.58$\pm$1.72 & 10328.12$\pm$1.22 & 548.31$\pm$1.51 &  4.497$\pm$0.012\\
GJ~1      &  5634.68$\pm$0.86 & -2337.71$\pm$0.71 & 230.42$\pm$0.90 &  3.711$\pm$0.015\\
GJ~27.1   &   485.57$\pm$2.81 &  -223.13$\pm$2.11 &  41.69$\pm$2.80 &  0.157$\pm$0.011\\
GJ~118    &   978.61$\pm$2.27 &   633.21$\pm$2.64 &  85.87$\pm$1.99 &  0.364$\pm$0.008\\
GJ~160.2  &    51.90$\pm$1.24 &  -780.04$\pm$1.57 &  43.25$\pm$1.61 &  0.325$\pm$0.012\\
GJ~173    &  -225.75$\pm$1.94 &  -192.57$\pm$1.88 &  90.10$\pm$1.74 &  0.022$\pm$0.000\\
GJ~180    &   408.07$\pm$2.49 &  -642.82$\pm$2.06 &  82.52$\pm$2.40 &  0.161$\pm$0.005\\
GJ~190    &   502.99$\pm$1.32 & -1399.76$\pm$1.52 & 107.85$\pm$2.10 &  0.471$\pm$0.009\\
GJ~218    &   768.13$\pm$1.39 &  -123.17$\pm$1.54 &  66.54$\pm$1.43 &  0.209$\pm$0.005\\
GJ~229    &  -137.09$\pm$0.50 &  -713.66$\pm$0.81 & 173.81$\pm$0.99 &  0.070$\pm$0.000\\
GJ~263    &  -123.73$\pm$3.68 &  -814.92$\pm$2.87 &  62.41$\pm$3.16 &  0.250$\pm$0.013\\
GJ~357    &   136.67$\pm$1.53 &  -989.13$\pm$1.41 & 110.82$\pm$1.92 &  0.207$\pm$0.004\\
GJ~377    & -1096.84$\pm$2.11 &   647.29$\pm$2.63 &  61.39$\pm$2.55 &  0.607$\pm$0.025\\
GJ~422    & -2466.98$\pm$2.87 &  1180.09$\pm$2.17 &  78.91$\pm$2.60 &  2.178$\pm$0.072\\
GJ~433    &   -72.51$\pm$1.49 &  -851.92$\pm$0.88 & 112.58$\pm$1.44 &  0.149$\pm$0.002\\
GJ~477    &  -101.39$\pm$2.28 &  -697.57$\pm$1.75 &  52.67$\pm$3.05 &  0.217$\pm$0.013\\
GJ~510    &  -452.04$\pm$2.55 &  -104.79$\pm$1.55 &  59.72$\pm$2.43 &  0.083$\pm$0.003\\
GJ~620    &  -348.40$\pm$2.35 &  -675.73$\pm$1.87 &  60.83$\pm$2.06 &  0.218$\pm$0.007\\
GJ~637    &  -480.59$\pm$1.17 &  -529.12$\pm$1.86 &  62.97$\pm$1.99 &  0.186$\pm$0.006\\
GJ~682    &  -708.98$\pm$2.55 &  -937.40$\pm$1.88 & 196.90$\pm$2.15 &  0.161$\pm$0.002\\
GJ~739    &   153.46$\pm$2.91 &  -495.17$\pm$1.98 &  70.95$\pm$2.56 &  0.087$\pm$0.003\\
GJ~817    &  -918.58$\pm$3.68 & -2038.13$\pm$2.50 &  52.16$\pm$2.92 &  2.202$\pm$0.123\\
GJ~821    &   713.47$\pm$2.82 & -1994.64$\pm$0.95 &  82.18$\pm$2.17 &  1.255$\pm$0.033\\
GJ~842    &   888.07$\pm$1.31 &  -125.56$\pm$1.16 &  83.43$\pm$1.77 &  0.222$\pm$0.005\\
GJ~855    &   587.72$\pm$2.17 &  -376.75$\pm$1.32 &  52.22$\pm$2.17 &  0.214$\pm$0.009\\
GJ~891    &   717.43$\pm$3.40 &    22.12$\pm$2.69 &  62.17$\pm$3.27 &  0.190$\pm$0.010\\
GJ~911    &   -42.36$\pm$3.32 &   128.87$\pm$2.44 &  41.22$\pm$2.64 &  0.010$\pm$0.001\\
GJ~1009   &    62.95$\pm$2.45 &  -196.98$\pm$2.46 &  55.62$\pm$2.32 &  0.018$\pm$0.001\\
GJ~1046   &  1395.67$\pm$1.72 &   547.16$\pm$2.52 &  71.06$\pm$3.23 &  0.727$\pm$0.033\\
GJ~1100   &   118.29$\pm$2.67 &  -498.47$\pm$1.30 &  34.57$\pm$2.79 &  0.174$\pm$0.014\\
GJ~3020   &   -44.73$\pm$4.05 &  -237.17$\pm$3.70 &  43.89$\pm$4.39 &  0.030$\pm$0.003\\
GJ~3082   &   104.14$\pm$2.04 &   311.53$\pm$1.81 &  60.38$\pm$1.81 &  0.041$\pm$0.001\\
GJ~3098   &  -589.19$\pm$1.86 &  -887.83$\pm$1.24 &  55.98$\pm$1.91 &  0.466$\pm$0.016\\
GJ~3671   &  -603.54$\pm$1.64 &  -296.43$\pm$1.36 &  56.38$\pm$2.04 &  0.184$\pm$0.007\\
GJ~3759   &  -391.22$\pm$1.50 &  -411.01$\pm$1.71 &  58.94$\pm$2.40 &  0.126$\pm$0.005\\
GJ~3916   &  -332.19$\pm$2.90 &  -352.83$\pm$2.62 &  66.21$\pm$3.18 &  0.082$\pm$0.004\\
GJ~3973   &    -9.48$\pm$2.45 &  -221.00$\pm$1.72 &  54.86$\pm$2.18 &  0.020$\pm$0.001\\
GJ~4106   &    31.55$\pm$4.63 &  -104.82$\pm$3.59 &   9.05$\pm$3.70 &  0.030$\pm$0.012\\
GJ~4293   &   198.08$\pm$2.42 &  -113.57$\pm$2.08 &  39.90$\pm$3.04 &  0.030$\pm$0.002\\
HG~7-15   &   176.02$\pm$2.85 &     5.73$\pm$1.79 &  26.80$\pm$2.05 &  0.027$\pm$0.002\\
Prox~Cen  & -3775.75$\pm$1.63 &   765.54$\pm$2.01 & 771.64$\pm$2.60 &  0.442$\pm$0.002\\
\hline
GJ~451~A  &  4003.98$\pm$0.37 & -5813.62$\pm$0.23 & 109.99$\pm$0.41 & 10.411$\pm$0.039\\
GJ~9511~B &  -999.75$\pm$1.29 & -3542.60$\pm$1.13 &  35.14$\pm$1.48 &  8.861$\pm$0.373\\
GJ~9511~A &  -997.47$\pm$1.20 & -3543.55$\pm$1.03 &  34.65$\pm$1.28 &  8.988$\pm$0.332\\
GJ~191    &  6505.08$\pm$0.98 & -5730.84$\pm$0.96 & 255.66$\pm$0.91 &  6.756$\pm$0.024\\
GJ~9371   &   264.99$\pm$2.23 & -3157.36$\pm$2.30 &  42.79$\pm$2.70 &  5.391$\pm$0.340\\
\hline
\end{tabular}
\end{table*}

Table~\ref{Tab:SecuAcc} lists the prediction for secular acceleration for all of our sample stars as derived from Eq.~(\ref{eq:SecuAcc}). Some stars (e.g. Barnard's star, GJ~1, and Proxima~Cen) have such high proper motion that the secular acceleration can be measured (4.5, 3.7, and 0.4\,m/s/yr, respectively). This effect is depicted with a dashed line for Proxima~Cen and for all stars with $\dot{v}_{\mathrm{r}}>1\,\mathrm{m/s/yr}$ in Fig.~\ref{Fig:RV1}. The secular acceleration was subtracted before subsequent analysis of the RV data.

We note that, even though Barnard's star has the highest (angular) proper motion $\mu$ and the dependence is quadratic on $\mu$, it is \emph{not} the star with the highest secular acceleration. When inspecting high proper motion stars from the Hipparcos catalogue \citep{vanLeeuwen07} we found four stars that have a higher secular acceleration because of their smaller parallax (see lower part of Table~\ref{Tab:SecuAcc}), namely: GJ~451 (Groombridge 1830, G8Vp + M5.5V), GJ~9511 (K2V + K2Vfe), GJ~191 (Kapteyn's star, M1V), and GJ~9371 (sdK4).

\subsection{Tests for variability and trends}

Following the recipe outlined by \citet{Endl02}, we performed several statistical tests to identify variability and RV trends in our data. First, we asked the question for each star of whether the observed variability or rms $\sigma$ is significantly higher than the mean measurement error $\overline{\sigma_{RV}}$ using the $F$-Test (and $F=\frac{\sigma^{2}}{\overline{\sigma_{RV}}^{2}}$ as $F$-value)\footnote{We use the one-tailed $F$-test because we are not interested in cases of error overestimation.}.

\begin{table*}
\caption{\label{Tab:StatTests}Tests for variability and trends. Listed are the number of measurements $N$, time baseline $T$, RV scatter $\sigma$ (rms), mean RV measurement error $\overline{\sigma_{RV}}$, and results of statistical tests. Low values of the $F$-test probability $P(F)$, as well as $P(\chi_{\mathrm{const}}^{2})$, e.g. \textless{}0.01, mean that a constant model is improbable, hence indicate variability (values in bold fonts). In contrast, a high $P(\chi_{\mathrm{slope}}^{2})$ indicates that a trend is an acceptable model while a small $P(F_{\mathrm{slope}})$ indicates a significant fit improvement.}
\centering
\begin{tabular}{lrrrrlclcllll}
\hline
\hline Star & $N$ &
\multicolumn{1}{c}{$T$}&
\multicolumn{1}{c}{$\sigma$}&
\multicolumn{1}{c}{$\overline{\sigma_{RV}}$}&
\multicolumn{1}{c}{$P(F)$}&
\multicolumn{1}{c}{$\chi_{\mathrm{const}}^{2}$}&
\multicolumn{1}{c}{$P(\chi_{\mathrm{const}}^{2})$}&
\multicolumn{1}{l}{Slope}&
$\chi_{\mathrm{slope}}^{2}$&
\multicolumn{1}{c}{$P(\chi_{\mathrm{slope}}^{2})$}&
$P(F_{\mathrm{slope}})$&
Comment\\
& & \multicolumn{1}{c}{{[}d{]}} & {[}m/s{]} & {[}m/s{]} & & & & {[}m/s/yr{]} & & & & \\
\hline
Barnard  & 75 & 2358 & 3.3 & 2.7 & 0.065 & 114 & \bf{0.0022} & -0.688 & 99.4 & \bf{0.022} & \bf{0.0038}\\
GJ~1     & 24 & 2151 & 2.5 & 2.4 & 0.9 & 27.3 & 0.24 & -0.454 & 25.9 & \bf{0.26} & 0.56\\
GJ~27.1  & 30 & 2177 & 6.1 & 6.1 & 0.95 & 41.2 & 0.066 & 0.122 & 41.1 & \bf{0.052} & 0.32\\
GJ~118   & 26 & 2266 & 6.5 & 5.8 & 0.56 & 40.5 & 0.026 & 1.99 & 23.8 & \bf{0.47} & \bf{0.00083}\\
GJ~160.2 & 33 & 2325 & 8.1 & 7.7 & 0.79 & 41.1 & 0.13 & 1.36 & 35.5 & \bf{0.26} & 0.067\\
GJ~173   & 12 &  897 & 5.3 & 3.1 & 0.094 & 30.6 & \bf{0.0013} & 1.61 & 28.7 & 0.0014 & 0.86\\
GJ~180   & 24 & 2325 & 3.8 & 4.1 & 0.71 & 29.4 & 0.17 & 0.418 & 27.5 & \bf{0.19} & 0.46\\
GJ~190   &  2 &    4 & 861.8 & 80.7 & 0.12 & 97.4 & \bf{$\bf{<10^{-7}}$} & -1.12$\cdot10^{5}$ & 0 & $<10^{-7}$ & \bf{$\bf{<10^{-7}}$} & SB2\\
GJ~218   &  9 &  896 & 3.1 & 3.1 & 0.98 & 8.49 & 0.39 & 1.03 & 7.96 & \bf{0.34} & 0.96\\
GJ~229   & 32 & 2325 & 5.5 & 2.8 & \bf{0.00036} & 139 & \bf{$\bf{<10^{-7}}$} & 1.43 & 95.6 & $<10^{-7}$ & \bf{0.0017}\\
GJ~263   &  2 &   82 & 117.6 & 78.6 & 0.75 & 1.19 & 0.27 & 743 & 0 & $<10^{-7}$ & \bf{$\bf{<10^{-7}}$} & SB2\\
GJ~357   & 30 & 2321 & 5.3 & 3.2 & \bf{0.0096} & 59.9 & \bf{0.00064} & 0.393 & 57.5 & 0.00084 & 0.58\\
GJ~377   & 14 & 1089 & 6.7 & 3.2 & 0.014 & 40 & \bf{0.00014} & -0.284 & 39.8 & 7.7$\cdot10^{-5}$ & 0.36\\
GJ~422   & 15 & 1112 & 4.0 & 3.4 & 0.55 & 16.8 & 0.27 & 0.599 & 16.2 & \bf{0.24} & 0.99\\
GJ~433   & 54 & 2554 & 4.4 & 3.6 & 0.16 & 80 & \bf{0.0097} & 0.284 & 78.4 & \bf{0.011} & 0.61\\
GJ~477   &  8 &  389 & 3486.0 & 4.3 & $\bf{1.6\cdot10^{-19}}$ & 5.58$\cdot10^{6}$ & \bf{$\bf{<10^{-7}}$} & 6.99$\cdot10^{3}$ & 4.52$\cdot10^{5}$ & $<10^{-7}$ & \bf{0.00034} & SB1\\
GJ~510   & 23 & 1115 & 5.6 & 3.5 & 0.039 & 54 & \bf{0.00016} & 2.64 & 37.9 & \bf{0.013} & 0.014\\
GJ~620   &  5 &  422 & 7.3 & 4.4 & 0.36 & 13.1 & 0.011 & -5.1 & 11 & \bf{0.012} & 0.98\\
GJ~637   & 17 & 1099 & 6.4 & 4.4 & 0.16 & 27 & 0.041 & 0.305 & 26.9 & \bf{0.03} & 0.38\\
GJ~682   & 20 & 1134 & 4.0 & 2.3 & 0.024 & 53.8 & $\bf{3.6\cdot10^{-5}}$ & 1.62 & 40.8 & 0.0016 & 0.056\\
GJ~739   & 19 & 1070 & 4.4 & 3.1 & 0.13 & 48.8 & \bf{0.00012} & -2.09 & 37.4 & 0.003 & 0.072\\
GJ~817   & 25 & 1551 & 4.9 & 4.3 & 0.54 & 32.2 & 0.12 & 0.184 & 32.1 & \bf{0.098} & 0.42\\
GJ~821   & 35 & 1516 & 5.0 & 3.8 & 0.12 & 53.7 & 0.017 & -0.268 & 53.5 & \bf{0.013} & 0.56\\
GJ~842   & 17 &  926 & 6.7 & 4.2 & 0.065 & 44.6 & \bf{0.00016} & -1.2 & 43.5 & 0.00013 & 0.91\\
GJ~855   & 22 & 1561 & 5.8 & 4.5 & 0.24 & 28.7 & 0.12 & -1.14 & 24.9 & \bf{0.21} & 0.19\\
GJ~891   & 25 & 2178 & 7.5 & 5.1 & 0.068 & 48.4 & \bf{0.0023} & 0.556 & 47.6 & 0.0019 & 0.93\\
GJ~911   & 17 & 2136 & 14.9 & 7.7 & 0.012 & 25.2 & 0.067 & 0.389 & 25.1 & \bf{0.049} & 0.38\\
GJ~1009  & 22 & 2177 & 5.3 & 4.0 & 0.23 & 47.8 & \bf{0.00074} & 0.0316 & 47.8 & 0.00046 & 0.055\\
GJ~1046  & 14 &  766 & 1248.3 & 3.6 & $\bf{1.9\cdot10^{-30}}$ & 1.59$\cdot10^{6}$ & \bf{$\bf{<10^{-7}}$} & 1.09$\cdot10^{3}$ & 8.8$\cdot10^{5}$ & $<10^{-7}$ & 0.018 & SB1~(BD)\\
GJ~1100  & 12 &  897 & 9.3 & 5.1 & 0.061 & 37 & \bf{0.00011} & -1.67 & 36.3 & 7.4$\cdot10^{-5}$ & 0.66\\
GJ~3020  & 13 &  749 & 298.8 & 9.0 & $\bf{5\cdot10^{-16}}$ & 7.62$\cdot10^{3}$ & \bf{$\bf{<10^{-7}}$} & -287 & 2.29$\cdot10^{3}$ & $<10^{-7}$ & \bf{0.00074} & SB1\\
GJ~3082  & 10 &  761 & 6.2 & 4.2 & 0.27 & 17.5 & 0.041 & 1.19 & 17.1 & \bf{0.029} & 0.66\\
GJ~3098  &  9 &  733 & 9.1 & 4.7 & 0.079 & 26.7 & \bf{0.00081} & 7.05 & 16 & \bf{0.025} & 0.13\\
GJ~3671  & 12 & 1090 & 5.6 & 4.4 & 0.46 & 17.4 & 0.095 & 3.26 & 10.2 & \bf{0.43} & 0.046\\
GJ~3759  & 11 & 1080 & 3.9 & 3.6 & 0.81 & 11.5 & 0.32 & 2.48 & 6.84 & \bf{0.65} & 0.071\\
GJ~3916  &  6 &  406 & 2170.7 & 9.2 & $\bf{1.5\cdot10^{-11}}$ & 4.07$\cdot10^{5}$ & \bf{$\bf{<10^{-7}}$} & 4.1$\cdot10^{3}$ & 2.12$\cdot10^{3}$ & $<10^{-7}$ & $\bf{2\cdot10^{-5}}$ & SB1\\
GJ~3973  &  5 &  420 & 6.8 & 3.6 & 0.25 & 12 & 0.017 & 3.3 & 10.8 & \bf{0.013} & 0.78\\
GJ~4106  &  5 &  396 & 20.7 & 15.9 & 0.62 & 7.4 & 0.12 & 32.2 & 1.99 & \bf{0.58} & 0.13 & giant\\
GJ~4293  & 14 &  875 & 8.7 & 5.6 & 0.13 & 32.3 & \bf{0.0021} & -0.114 & 32.3 & 0.0012 & 0.064\\
HG~7-15  & 11 &  417 & 8.7 & 10.3 & 0.59 & 11 & 0.36 & 10.5 & 9.03 & \bf{0.43} & 0.39\\
Prox~Cen & 76 & 2555 & 3.6 & 2.3 & \bf{0.00028} & 183 & \bf{$\bf{<10^{-7}}$} & 0.703 & 159 & $<10^{-7}$ & \bf{0.0026}\\
\hline
\end{tabular}
\end{table*}

The calculated probabilities $P(F)$ from the $F$-Test are listed in Table~\ref{Tab:StatTests} for each star. A low value of $P(F)$ (e.g. $<0.01$, i.e. 99\% confidence, in bold face in Table~\ref{Tab:StatTests}) indicates that the observed scatter can probably not be explained with the measurement errors and that there is an excess variability or a trend. This is the case for the stars with a high rms (GJ~477, GJ~1046, GJ~3020, and GJ~3916), which seem to have a companion, probably in the brown dwarf or low-mass star regime (see also Fig.~\ref{Fig:RV2}). GJ~190 and GJ~263 also have a high sample variance. They may have a stellar companion bright enough to contaminate the spectrum. We deduce this from the large measurement error that would occur for a double-lined spectroscopic binary (SB2) because our data modelling is only designed for a single-lined spectroscopic binary (SB1). Also, because there are only two measurements for GJ~190 and GJ~263, they do not stand out in the $F$-statistics. Indeed GJ~263 has already been identified as a spectroscopic binary, and an adaptive optics image was presented by \citet{Beuzit04}. Among the stars with an rms smaller than 20\,m/s, only GJ~229, GJ~357, and Proxima Cen have an rms that is significantly greater than the measurement error.

A similar test for RV variability is to determine the goodness of fit for a constant model, i.e. calculating the $\chi^{2}$ above the weighted RV mean\footnote{When each measurement has the same error $\sigma_{RV}$, one gets
\[
\chi_{\mathrm{red}}^{2}=\frac{\chi_{\mathrm{const}}^{2}}{N-1}=\frac{1}{(N-1)\sigma_{RV}^{2}}\sum(RV_{i}-\overline{RV})^{2}=\frac{\sigma^{2}}{\sigma_{RV}^{2}}=F.
\]
} and deriving the probability from the $\chi^{2}$-distribution. These probability values $P(\chi_{\mathrm{const}}^{2})$ are mostly lower than $P(F)$, and therefore there would be more variable stars according to our $P<0.01$ criterion (Table~\ref{Tab:StatTests}).

As long-period planets can cause a trend in the RV, we also tested for this by weighted fitting of a linear slope. A high probability of the resulting $\chi_{\mathrm{slope}}^{2}$ indicates that this is an acceptable fit (on the contrary, a low probability indicates remaining variability). But it is also informative to compare $\chi_{\mathrm{slope}}^{2}$ with the $\chi_{\mathrm{const}}^{2}$ of the weighted mean via the $F$-value \citep{Cumming99}
\[
F_{\mathrm{slope}}=(N-2)\frac{\chi_{\mathrm{constant}}^{2}-\chi_{\mathrm{slope}}^{2}}{\chi_{\mathrm{slope}}^{2}}.
\]
A high $F_{\mathrm{slope}}$-value indicates a fit improvement whereas a low probability $P(F_{\mathrm{slope}})$ shows this improvement to be significant and that it is probable not due to noise.

In Table~\ref{Tab:StatTests} the stars with a significant trend are marked in bold face. Besides the companion hosting M~dwarfs GJ~477, GJ~3020, and GJ~3916, these are the stars Barnard's star, GJ~118, GJ~229, and Proxima~Cen. The significant trend of 1.4~m/s/yr for GJ~229 may be caused by the wide T7p brown dwarf companion GJ~229~B, as one can show with a rough estimate. Assuming a circular orbit and the projected separation of 45\,AU ($\alpha=7.7\arcsec$ \citealt{Nakajima95}, $\pi=0.173\arcsec$) as the orbital radius to GJ~229, the orbital period is of the order of $P=400\,$yr. Then a 30$\, M_{\mathrm{Jup}}$ companion can cause a velocity amplitude of $K=170\,$m/s (Eq.~(\ref{eq:Amplitude})) and the maximum RV change is $\dot{RV}=K\frac{2\pi}{P}=2.7\,\mathrm{m/s/yr}$, about twice as large as the observed trend. However, the low $P(\chi_{\mathrm{slope}}^{2})$ indicates there is still some variability after trend subtraction.

\subsection{Periodogram analysis}

To test for periodicities in the RV data, we computed the generalized Lomb-Scargle periodogram (GLS, \citealt{Zechmeister09}), which is the equivalent of fitting sine waves including an offset. The adopted period search interval ranges from 2\,d to the time baseline $T$ of each data set. Note that 2\,d will mostly exceed the average Nyquist frequency. However, searching at frequencies higher than the Nyquist frequency is possible for irregular sampling \citep{Pelt09}. False-alarm probabilities (FAP) were determined by bootstrap randomization \citep[e.g.][]{Kuerster97}. In this method random data sets are generated from an original data set by shuffling the RVs while retaining the observing times. For each random data set, the GLS periodogram was computed and searched for its maximum. The FAP is then given by that fraction of random data sets having a periodogram power higher than the original one. For each star we generated 1000 random data sets which resolves FAP~$>10^{-3}$ and is sufficient to decide whether the FAP is below our threshold of 0.01.

Table~\ref{Tab:Periodicites} shows that GJ~4106 and GJ~1046 have a FAP marginally lower than 0.01. The probable brown dwarf companion to GJ~1046 with a minimum mass of 26.9$\, M_{\mathrm{Jup}}$ in an eccentric orbit ($e=0.28$) with a 168.8\,d period has been already published by \citet{Kuerster08} based on this UVES data set. For the 365\,d period in Proxima Cen we refer to \citet{Endl08b}, who recently analysed the RV data and identified this period as a 1-year alias.

\begin{figure*}
\includegraphics{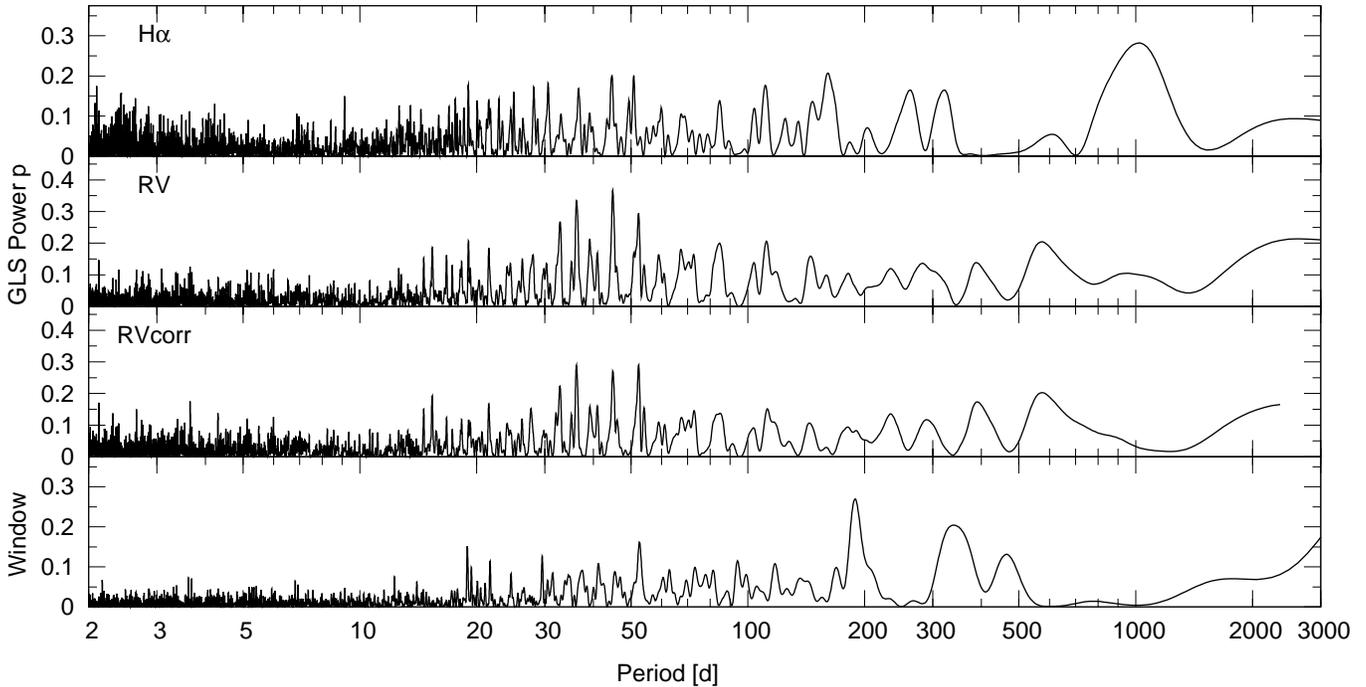}
\caption{\label{Fig:GLS}GLS periodograms for the H$\alpha$-index (top panel; see Sect.~\ref{sub:Correlation-HRV}), the RV data (second panel), the RV data corrected for a H$\alpha$ correlation (third panel), and window function (bottom) of Barnard's star.}
\end{figure*}

\begin{figure}
\includegraphics{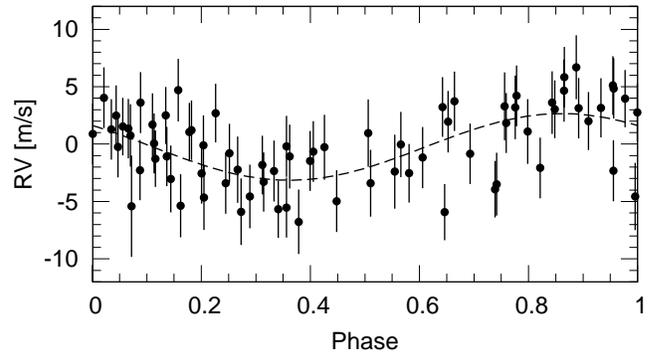}
\caption{\label{Fig:Barnardphased}Radial velocity time series for Barnard's star phased to the 44.9\,d period and the best-fitting sinusoid.}
\end{figure}

Based on data from the first 2\,1/2\,yr, \citet{Kuerster03} determined two RV periods of 32\,d and 45\,d for Barnard's star with a FAP of 0.56\% and 1.05\%, respectively. Now, in the enlarged data set with a 6\,1/2\,yr time baseline, the second period (45\,d; with an amplitude of 2.9\,m/s) has the highest periodogram peak, and its FAP is now less than 0.1\%, i.e. more significant (see Fig.~\ref{Fig:GLS}, second panel for the periodogram and Fig.~\ref{Fig:Barnardphased} for the RVs phased to this period). However, stellar activity is the probable cause of this variability and will be discussed in Sect.~\ref{sub:Correlation-HRV}.

\begin{table}
\caption{\label{Tab:Periodicites}Test for periodicities in the RV (GJ~190 and GJ~263 excluded): the best period $P$ with its $\chi^{2}$ and bootstrapped FAP.}
\centering
\begin{tabular}{lccl}
\hline
\hline Star & $P$ {[}d{]} & $\chi_{\mathrm{sin}}^{2}$ & \multicolumn{1}{c}{FAP}\\
\hline
Barnard  & 44.9 & 72.5 & $\mathbf{<10^{-4}}$\\
GJ~1     & 2.73 & 12.5 & 0.733\\
GJ~27.1  & 2.01 & 18   & 0.189\\
GJ~118   & 3.97 & 15.6 & 0.329\\
GJ~160.2 & 4.15 & 20.7 & 0.577\\
GJ~173   & 2.28 & 3.67 & 0.026\\
GJ~180   & 5.93 & 8.63 & 0.127\\
GJ~218   & 2.02 & 0.473 & 0.557\\
GJ~229   & 10.9 & 63.7 & 0.102\\
GJ~357   & 3.41 & 28.4 & 0.157\\
GJ~377   & 15.1 & 7.47 & 0.029\\
GJ~422   & 8.82 & 4.69 & 0.498\\
GJ~433   & 6.5  & 55.6 & 0.451\\
GJ~477   & 243  & 1.58\,$\cdot10^{5}$ & 0.541\\
GJ~510   & 2.92 & 29.1 & 0.867\\
GJ~620   & 2.51 & 0.00667 & 0.637\\
GJ~637   & 8.54 & 8.54 & 0.245\\
GJ~682   & 89.3 & 20.1 & 0.178\\
GJ~739   & 2.34 & 16.3 & 0.334\\
GJ~817   & 2.36 & 15.1 & 0.743\\
GJ~821   & 12.6 & 32.9 & 0.393\\
GJ~842   & 92.6 & 10.8 & 0.219\\
GJ~855   & 16   & 12.6 & 0.668\\
GJ~891   & 30.5 & 22.1 & 0.36\\
GJ~911   & 2.35 & 9.85 & 0.909\\
GJ~1009  & 3.73 & 17.1 & 0.502\\
GJ~1046  & 174  & 1.23\,$\cdot10^{5}$ & \bf{0.008}\\
GJ~1100  & 3.79 & 5.27 & 0.477\\
GJ~3020  & 5.14 & 3.03\,$\cdot10^{3}$ & 0.895\\
GJ~3082  & 3.72 & 1.7  & 0.261\\
GJ~3098  & 2.7  & 2.23 & 0.5\\
GJ~3671  & 68.1 & 1.69 & 0.218\\
GJ~3759  & 2.46 & 1.8  & 0.769\\
GJ~3916  & 8.6  & 82.5 & 0.077\\
GJ~3973  & 2.25 & 0.00154 & 0.105\\
GJ~4106  & 2.41 & 0.000139 & \bf{0.009}\\
GJ~4293  & 19.1 & 11    & 0.994\\
HG~7-15  & 4.55 & 0.919 & 0.179\\
Prox~Cen & 365  & 101   & $\mathbf{<10^{-4}}$\\
\hline
\end{tabular}
\end{table}

\subsection{\label{sub:Upper-detection-limits}Upper detection limits}

A Jupiter-mass planet in a circular orbit with a radius of 1\,AU around a $0.4M_{\odot}$-M~dwarf would cause an RV amplitude of 45\,m/s (even higher for closer orbits; Eq.~(\ref{eq:Amplitude})) and would be easily detectable with our precision of typically a few m/s. However, most of our sample stars show low RV-variations and no indication of a planet. To determine which planets in circular orbits can be excluded, we calculated detection limits in the following way.

\begin{figure*}
\centering
\includegraphics[height=0.96\textheight,width=1\linewidth]{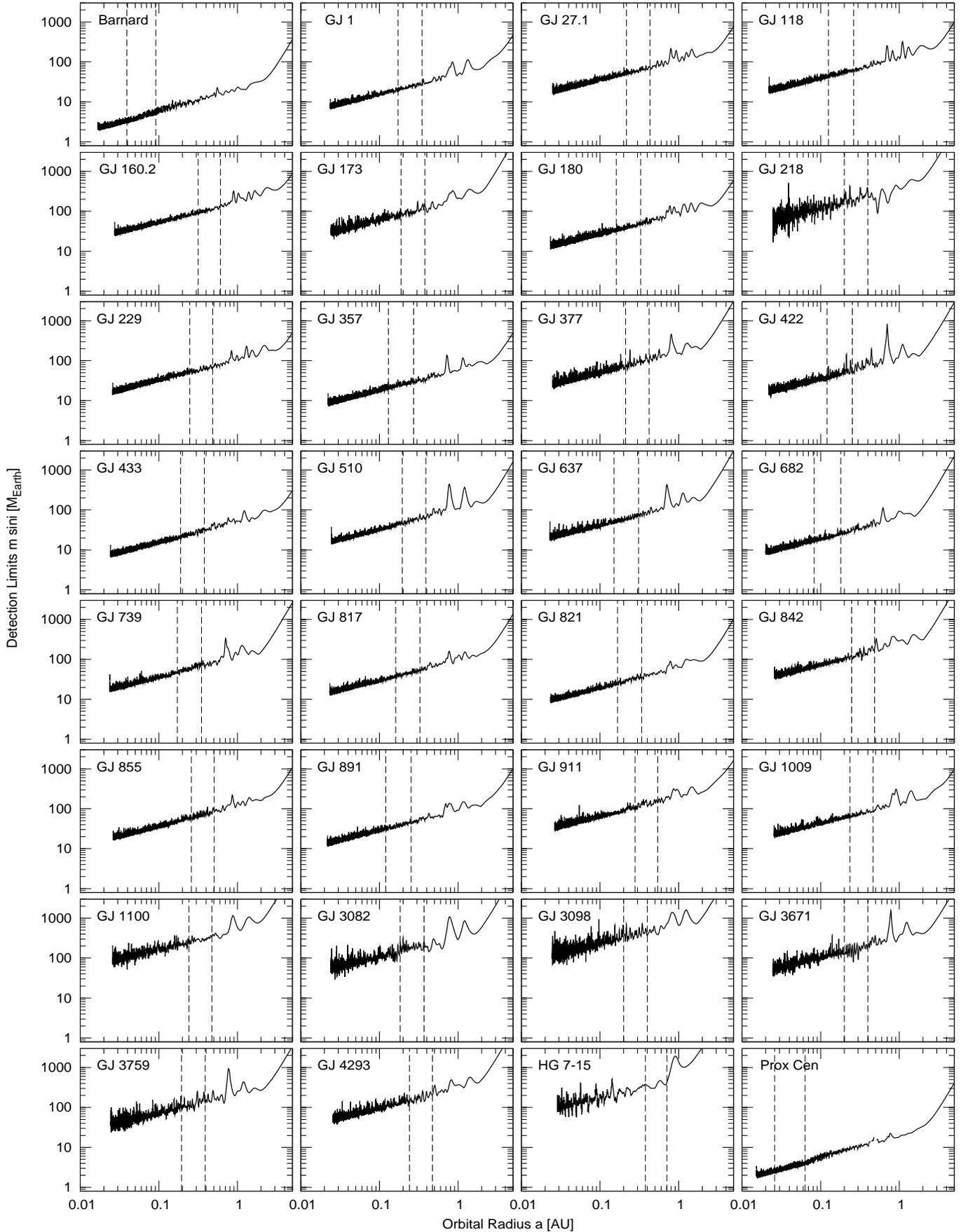}
\caption{\label{Fig:DetectLimits}Detection limits for stars with at least 8 RV measurements. Planets in circular orbits above the limit would be detected with a probability \textgreater{}99\%. The dashed lines show the habitable zone (HZ) \citep{Kasting93}.}
\end{figure*}

We considered the data as noise and simulated planetary signals by adding sine waves to the data with a range of trial frequencies and for 12 different equidistant phases. The time sampling remained unaffected. For the simulated data the generalised Lomb-Scargle periodogram power was calculated at the trial sine wave frequency where the peak of the signal is expected. If this power was below a power threshold (which corresponds to an FAP of 0.01, see note below), we increased the amplitude of the sine wave. The simulated planet is considered as detected, if the power is in all 12 phases (equal to or) higher than the threshold. The corresponding amplitude is considered as the 99\% detection limit. In Appendix~\ref{sec:Amp_limits} we provide an analytic solution to calculate this amplitude for a given power threshold and phase. This procedure was performed for a number of trial frequencies resulting in amplitude limits which finally were translated into mass upper limits (Fig.~\ref{Fig:DetectLimits}) using Eq.~(\ref{eq:Amplitude}) and the stellar masses from Table~\ref{Tab:Sample}.

It should be noted that the periodogram power $p$ is a measure/quantity for the FAP. By applying bootstrap randomisation to the original data we get a FAP vs. power relation $p(FAP)$. No assumptions are necessary about the number of independent frequencies. The power value for FAP=0.01 was also taken as the threshold for the simulated data. This modification to the method by \citet{Endl02} bypasses bootstrapping the simulated data again and is therefore more efficient and allows a dense sampling of the frequency.

Here we assume that the $p(FAP)$ relation does not change much when adding a sine wave, because the time sampling and number of measurements does not change. The only thing that can happen is that the rms (or $\chi^{2}$ above the mean) of the simulated data changes (increases) by a certain factor. But as the normalized power is invariant when the measurements are scaled the resulting effect is small. A comparison of the detections limits given in Fig.~\ref{Fig:DetectLimits} with those from \citet{Endl08b} for Proxima~Cen shows quite similar results. Figure~\ref{Fig:DetectLimits} also shows the steep increase that generally appears for periods longer than the time base (see also \citealt{Cumming04,Nelson98}).

Instead of again searching the whole frequency range, only the power at the original sine frequency was calculated, since one can expect to find the simulated signal there. This saves computational effort and is more conservative, because we exclude spurious detections that exceed the power threshold with a lower amplitude at alias or noise frequencies.

Figure~\ref{Fig:DetectLimits} shows the results of our detection limit calculation. It can be seen that for Barnard's star or Proxima~Cen, i.e. stars with low masses and many measurements, the detection limit reaches down to a few Earth masses for close-in circular orbits and even within their habitable zones (HZ). Both stars have a priori frequencies with a FAP\textless{}0.01. These frequencies were excluded from the detection limit calculation (Barnard's star: 36.1 -- 36.4\,d and 44.6 -- 45.1\,d; Proxima~Cen: 295.0 -- 313.11\,d and 347.1 -- 392.5\,d). The HZ is depicted for each star derived from Fig.~15 of \citet{Kasting93}, whereas we used the stellar masses from Table~\ref{Tab:Sample}. For an M~dwarf with a mass of 0.3$M_{\odot}$, the HZ is beyond 0.1\,AU. For several stars we can exclude planets with a few ten Earth masses in their HZ and Jupiter-mass planets ($1M_{\mathrm{Jup}}=318M_{\mathrm{Earth}}$) up to a few AU.

\subsection{\label{sub:Correlation-HRV}Correlation between RV and H$\alpha$~index?}

Stellar activity can affect the measured RV. The H$\alpha$~line is an indicator of stellar activity (the only available one in the UVES spectra). In Proxima~Cen as an active flare star, the H$\alpha$~line is in emission in contrast to low-activity Barnard's star. However, the H$\alpha$~line is variable in both cases. Therefore investigating the correlation between RV and variations in the H$\alpha$~line may be useful for correcting RV for stellar activity.

\citet{Kuerster03} report a correlation of this type for Barnard's star. We checked this again with the now available longer data set for Barnard's star, as well as for the other stars. As a measure of the variability of the H$\alpha$~line we adopt the definition of the H$\alpha$-index by \citet{Kuerster03}:
\begin{equation}
I=\frac{\overline{F_{0}}}{0.5(\overline{F_{1}}+\overline{F_{2}})}\label{eq:index}
\end{equation}
where $\overline{F_{0}}$ is the mean flux in the range of {[}-15.5km/s,+15.5km/s{]} around the H$\alpha$~line ($\lambda=656.28\,$nm) and $\overline{F_{1}}$ and $\overline{F_{2}}$ are the mean flux two reference bandpasses ({[}-700km/s,-300km/s{]} and {[}600~km/s, 1000~km/s{]}, respectively) used for normalization. This index is a kind of filling-in of the H$\alpha$~line (i.e for emission $I\ga1$) and is related to equivalent width (see Appendix~\ref{sec:index-EW}). Following \citet{Kuerster03}, we also computed a CaI index {[}+441.5~km/s, 472.5~km/s{]} for comparison. The CaI Line ($\lambda=6572.795\,$\AA) is expected to be stable.

\begin{figure}
\includegraphics{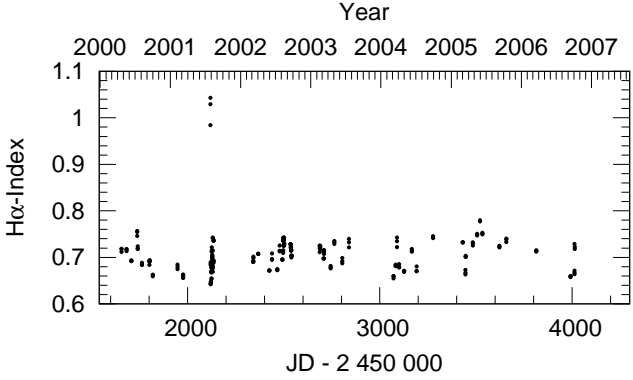}
\caption{\label{Fig:H-t_barnard}Time series of the H$\alpha$-index for Barnard's star.}
\end{figure}

\begin{figure}
\includegraphics{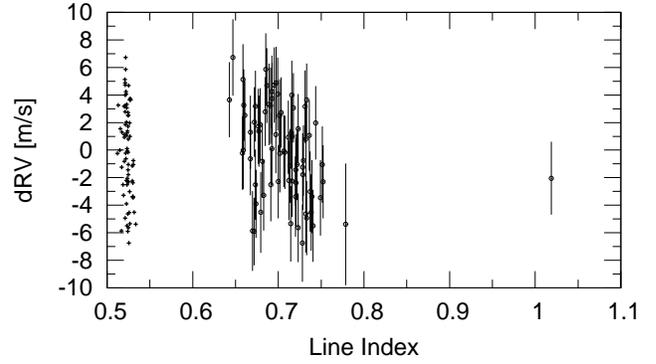}
\caption{\label{Fig:H-RV_barnard}Correlation of the H$\alpha$-index with the RV for Barnard's star (dots with error bars, $r_{\mathrm{H}\alpha}=0.42$). For comparison the CaI-index (crosses).}
\end{figure}

Figure~\ref{Fig:H-t_barnard} shows the variation in the H$\alpha$~index for Barnard's star with time. In Fig.~\ref{Fig:H-RV_barnard} the RV is plotted against the H$\alpha$~index. The flare event has already been described in \citet{Kuerster03}, does not seem to have any effect on the RV, and is excluded from the correlation analysis. No further flare was detected in the longer data set. We calculated a correlation coefficient of $r=-0.42$ for the whole data set. Because the new data (JD\textgreater{}2452600) have a less pronounced anti-correlation with a correlation coefficient of $r=-0.25$, the correlation decreases compared to the old data set where $r=-0.50$ \citep{Kuerster03}, but the anti-correlation is still present and becomes more significant due to the longer baseline.

\begin{figure}
\includegraphics{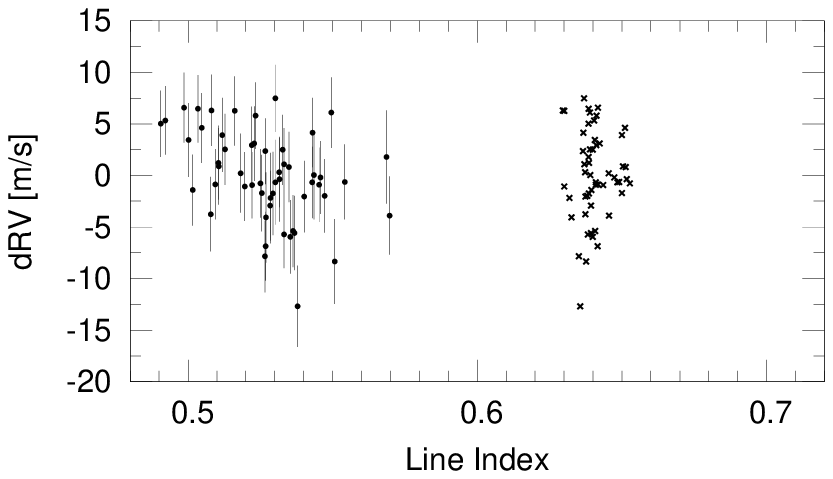}
\caption{\label{Fig:H-RV_gj433}Correlation of the H$\alpha$- and CaI-index with the RV for GJ~433 ($r_{\mathrm{H}\alpha}=-0.40$).}
\end{figure}

\begin{figure}
\includegraphics{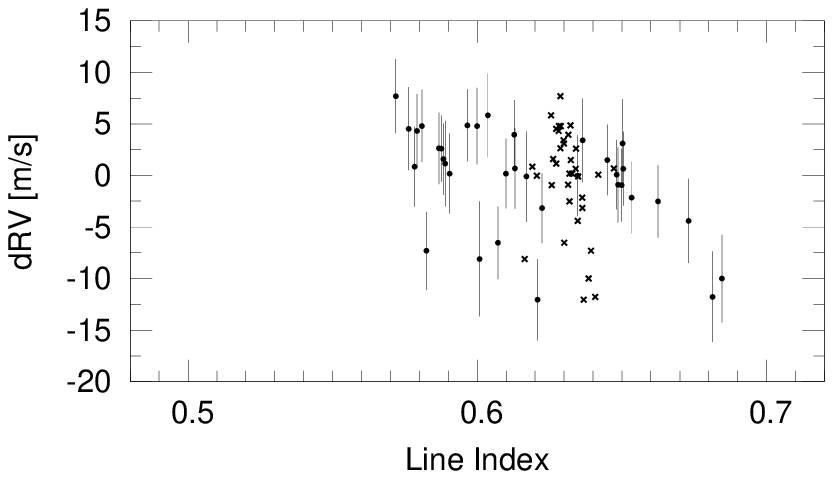}
\caption{\label{Fig:H-RV_gj821}Correlation of the H$\alpha$- and CaI-index with the RV for GJ~821 ($r_{\mathrm{H}\alpha}=-0.49$).}
\end{figure}

\begin{figure}
\includegraphics{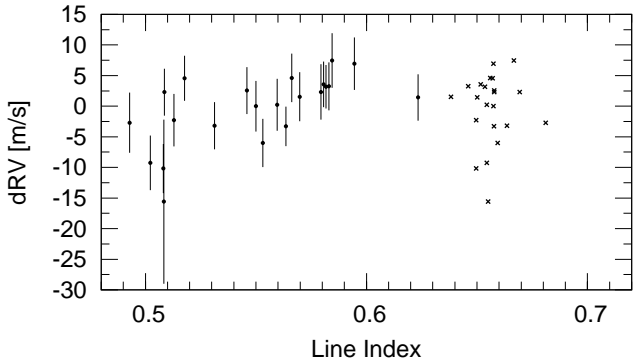}
\caption{\label{Fig:H-RV_gj855}Correlation of the H$\alpha$- and CaI-index with the RV for GJ~855 ($r_{\mathrm{H}\alpha}=0.62$).}
\end{figure}

We performed a period search (Fig.~\ref{Fig:GLS}, top panel) for the H$\alpha$~index of Barnard's star (again flare event is excluded). With the new data, a 1000\,d period dominates the GLS periodogram (upper panel). There is also some power at 44.5~d that was found to be the dominant period in the RV data (second panel). Therefore it is probable that the 45-day RV period is caused by stellar activity rather than by a planet (third panel). When we subtract the correlation from the RV data, the rms reduces slightly from 3.35~m/s to 3.09~m/s. The FAP of this period for the corrected RV data is only 5.5\% compared to $<0.01$\% of the uncorrected RV data (Table~\ref{Tab:Periodicites}). A similar analysis for the other stars\footnote{The 6 six stars with companions were excluded.} only yields a significant correlation for GJ~433, GJ~821, and GJ~855 (Figs.~\ref{Fig:H-RV_gj433}--\ref{Fig:H-RV_gj855}). All stars show no significant RV-CaI~index correlation.

\begin{figure}
\includegraphics{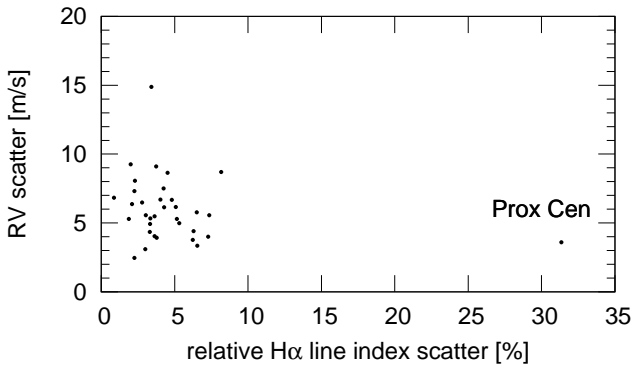}
\caption{\label{Fig:RV-H_scatter} Comparison of the RV scatter (rms) and H$\alpha$ variability for all M~dwarfs with an rms~$<20\,$m/s. No correlation can be seen. The very active star Proxima~Cen shows only very low RV excess.}
\end{figure}

Finally, we investigated the question of whether more active stars show more RV excess scatter. Figure~\ref{Fig:RV-H_scatter} compares the RV scatter (rms from Table~\ref{Tab:StatTests}) and the relative H$\alpha$ line index scatter for all M~dwarfs from our sample and demonstrates that no correlation can be seen.

\section{Discussion}

We browsed the available literature and catalogues for known binarity of our M dwarfs, which may have an impact on planet formation. Well known are the substellar companions to GJ~229 \citep{Nakajima95} and GJ~263 \citep{Beuzit04}. Proxima~Cen constitutes a widely separated common proper motion pair with $\alpha$~Cen A+B.

For GJ~477 and GJ~3916, binarity is indicated in the Hipparcos catalogue. GJ~477 has the double/multiple systems flag ``X'' (stochastic solution, probable astrometric binaries with a short period). GJ~3916 is listed with the flag ``G'' (acceleration or higher order terms). The large amplitudes seen in the RVs for both stars confirm this.

Other stars with an entry are: GJ~620 ({}``G''), GJ~4106 ({}``G''), and GJ~433 ({}``O'', i.e. orbital solution). For these stars we found only low RV variation. This is at least controversial for GJ~433 (HIP 56528). The announced period is 500\,d \cite[see also][]{Bernstein97} and the amplitude is expected to be several hundred m/s.\footnote{The parameters for the circular orbit ($e=0$, $\omega=0$) are $P=500\pm32\,$d, $T_{0}=2\,448\,402\pm28\,$d, $a_{0}=4.27\pm2.04\,$mas, $i=54\pm2$, and $\Omega=346\pm22\deg$. With the parallax $\pi=112.58\pm1.44$ the semi major axis of the photocentre is $a_{0}=0.0379\,$AU. Assuming $a_{0}\approx a_{1}$ the amplitude would be $K_{1}=\sin i\frac{2\pi a_{0}}{P}=668\,$m/s for the circular orbit. With $m_{1}=0.48M_{\odot}$ the secondary mass is $m_{2}=19.8M_{\mathrm{Jup}}$ calculated from the mass function $f(m)=\frac{(m\sin i)^{3}}{(m+M)^{2}}=\frac{P}{2\pi G}(K\sqrt{1-e^{2}})^{3}$, i.e. $m\sin i=K\cdot\sqrt[3]{(m+M)^{2}\frac{P}{2\pi G}}$.} At 1\,AU our detections limit reaches down to 0.2$M_{\mathrm{Jup}}$ (Fig.~\ref{Fig:DetectLimits}).

To our knowledge the rest of our sample so far has no discovered companions. Companions are explicitly excluded with near-infrared speckle interferometry by \citet{Leinert97} for GJ~1 and GJ~682 ($\Delta K=-4.5\,$mag at 1-10\,AU), as well as GJ~891 with infrared coronagraphic imaging by \citet{McCarthy04} (\textgreater{}30$M_{\mathrm{Jup}}$ at 140-1200AU).

\citet{Kuerster03} discussed how several stellar activity phenomena, such as spots, plages, or convective RV shifts, might affect the RV measurements. A linear RV-H$\alpha$ anti-correlation that is present in Barnard's star could be a result of a convective redshift caused by plages that suppress blueshifted convective flows. We find such an anticorrelation for GJ~433 and GJ~821, while GJ~855 exhibits a positive correlation (Figs.~\ref{Fig:H-RV_gj433}--\ref{Fig:H-RV_gj855}). \citet{Bonfils07} presented with GJ~674 (M2.5V) an example where the RV and H$\alpha$ variations seem to be phase-shifted. The correlation thus differs from a linear one and looks like a loop. One would expect such behaviour for a plage rotating with the surface.

Flare events, such as the one observed in Barnard's star, do not seem to take part in the correlation. This agrees with the fact that no significant and strong correlation is observed in the flaring M~dwarf Proxima~Cen \cite[see also][]{Endl08b}.

\section{Conclusion}

Within the sensitivity provided by our RV precision of a few m/s we have not detected any planets around our sample stars. Most of the M~dwarfs exhibit only low RV variations and some of them have a measurable secular acceleration due to their high proper motion.

We have discussed two effects on the RV that are not caused by planetary companions. First, the secular acceleration, is a perspective effect and causes always a positive RV trend. This effect can be corrected easily. Vise versa, this can be seen as an independent measurement of the astrometric quantity $\frac{\mu^{2}}{\pi}$, i.e. a confirmation for the ratio of squared proper motion and parallax (the absolute RV is not required). The second effect are RV variations caused by stellar activity. This is likely the case for the 45\,d period we found in the RV data of Barnard's star. The RVs correlate with the H$\alpha$ index. Such a correlation was found here only for a few M~dwarfs and therefore no conclusions can be drawn in general.

As a by-product of our survey we have identified 6 M~dwarfs with low-mass companions, four of them (GJ~477, GJ~1046 and GJ~3020, and GJ~3916) are brown dwarf or low-mass stellar candidates and two are spectroscopic binaries (SB2: GJ~263 and GJ~190). Follow-up RV observations will yield the orbital parameters and the lower limits for the companion masses $m\sin i$.

Our detection limits demonstrate that we can exclude giant planets with $1M_{\mathrm{Jup}}$ up to 1\,AU for half of our M~dwarfs and that we are in principle capable of discovering planets with a few Earth masses in the habitable zones of M~dwarfs with VLT+UVES. For this purpose an adequate number of measurements is needed to find low amplitudes of the order of the achieved high RV precision. The low frequency of Jupiter-mass planets around M~dwarfs requires a large sample. The given detection limits are based on the search for single planets in circular orbits. These limits would be higher for eccentric orbits or multi-planet systems.

Even if planet detections are more spectacular, it is also important to report non-detections, which are required to estimate the planet frequency. Our non-detections of planets support the increasing observational evidence of a lower frequency of Jupiter-mass planets around M dwarfs. \citet{Endl06} estimated a frequency of $\approx1$\% or less up to 1 AU orbital radius for Jupiter-mass planets around M~dwarfs compared to 2.5\% for solar like stars. This comparison cannot be done yet for low-mass planets because they are much harder to detect (and even more so for G~stars), which introduces observational biases. However low-mass planets seem to be quite frequent around M~dwarfs \citep{Bonfils07}.

Expansion of current M~dwarf planet searches will allow more precise determination of the true frequency of giant planetary companions to this type of stars and lower the detection limits. Photometric surveys for transits like the recently started project MEarth \citep{Irwin08} monitoring 2000 nearby M~dwarfs and microlensing projects will also contribute. The discoveries of planetary systems around GJ~876 \citep{Delfosse98,Marcy98,Marcy01,Rivera05} and GJ~581 \citep{Udry07} show that planets do exist around M~dwarfs. This promises further discoveries of low-mass planets in the future with high-precision RV surveys.

\begin{acknowledgements}
We thank the ESO OPC and the DDTC for the generous allocation of observing time. We are also grateful to all of ESO staff who helped with the preparation of the service mode observations, carried them out, or processed, verified, and distributed the data. A number of people have contributed in many ways to make this survey happen. Our thanks go to Artie P. Hatzes, St\'{e}phane Brillant, William D. Cochran, Sebastian Els, Thomas Henning, Andreas Kaufer, Sabine Reffert, Florian Rodler, Fr\'{e}d\'{e}ric Rouesnel, and Steve S. Saar. This material is based on work supported by the National Aeronautics and Space Administration under Grants NNG04G141G, NNG05G107G issued through the Terrestrial Planet Finder Foundation Science program and Grant NNX07AL70G issued through the Origins of Solar Systems Program.
\end{acknowledgements}

\bibliographystyle{aa}
\bibliography{paper}

\appendix

\section{Relation between index and equivalent width\label{sec:index-EW}}

The equivalent width is defined as
\[
EW=\sum_{i}\delta\lambda_{i}\frac{F_{C_{i}}-F_{i}}{F_{C_{i}}}
\]
where $\delta\lambda_{i}$ is the width of the $i$-th pixel in wavelength, and $F_{i}$ and $F_{C_{i}}$ are the flux and continuum flux in the $i$-th pixel, respectively. The equivalent width $EW$ is measured in terms of wavelengths. In a normalized spectrum the continuum flux is constant $F_{C_{i}}=F_{C}$ resulting in
\[
EW=\Delta\lambda\left[1-\frac{1}{\Delta\lambda}\sum\delta\lambda_{i}\frac{F_{i}}{F_{C}}\right]
\]
where $\Delta\lambda=\sum\delta\lambda_{i}$ is the considered wavelength range.

The index $I$ as defined (for H$\alpha$) in Eq.~(\ref{eq:index}) is, on the other hand, a dimensionless measure. It is normalized by the mean flux $\overline{F}=\frac{1}{\Delta\lambda}\sum_{i}\delta\lambda_{i}F_{i}$ taken from reference ranges instead of the continuum, which is sometimes difficult to estimate, in particular for M~dwarfs with their ubiquitous absorption lines.

In the case that the reference regions are estimated as continuum, i.e. $\overline{F_{1}}=\overline{F_{2}}=F_{C}$, the index becomes $I=\frac{1}{F_{C}}\frac{1}{\Delta\lambda}\sum_{i}\delta\lambda_{i}F_{i}$ and is related to $EW$ as
\[
I=1-\frac{EW}{\Delta\lambda}.
\]
Note that an absorption line is indicated by $EW>0$ ($0<I<1$) and an emission line by $EW<0$ ($I>1$).

\section{\label{sec:Amp_limits}Response of the GLS periodogram when adding a sine wave}

The definition of the generalized Lomb-Scargle periodogram (GLS) is (we use the notation introduced in \citealt{Zechmeister09})
\[
p_{y}(\omega)=\frac{1}{YY}\cdot\frac{SS\cdot YC^{2}+CC\cdot YS^{2}-2CS\cdot YC\cdot YS}{CC\cdot SS-CS^{2}}
\]
whereas abbreviations are weighted covariances for data $y_{i}$, sine, and cosine terms (e.g. $YC=\sum w_{i}y_{i}\cos\omega t_{i}-\sum w_{i}y_{i}\cdot\sum w_{i}\cos\omega t_{i}=\sum w_{i}(y_{i}-\overline{y})\cos\omega t_{i}$; $w_{i}$ are normalised weights).

When introducing a parameter $\tau$ defined by $\tan2\tau=\frac{2CS}{CC-SS}$ and replacing $t_{i}$ by $\tau_{i}=t_{i}-\tau$ (resulting in $CS_{\tau}=0$) the GLS can be written as (see \citet{Zechmeister09} for details)
\[
p_{y}(\omega)=\frac{1}{YY}\left[\frac{YC_{\tau}^{2}}{CC_{\tau}}+\frac{YS_{\tau}^{2}}{SS_{\tau}}\right]
\]
in a form very similar to classical Lomb-Scargle periodogram. While the first formulation can save some computational effort, the use of the second is more elegant for our purpose.

By adding a sine wave with frequency $\omega_{0}$ to the data $y_{i}$, we generate new data $x_{i}=y_{i}+a\cos\omega_{0}t_{i}+b\sin\omega_{0}t_{i}$ or $x_{i}=y_{i}+a_{\tau}\cos\omega_{0}\tau_{i}+b_{\tau}\sin\omega_{0}\tau_{i}$. This results in a new periodogram $p_{x}(\omega)$ for the new data set $x$
\begin{align*}
p_{x}(\omega) & =\frac{1}{XX}\left[\frac{XC_{\tau}^{2}}{CC_{\tau}}+\frac{XS_{\tau}^{2}}{SS_{\tau}}\right].
\end{align*}
Because the times are not changed, this neither affects the parameter $\tau$ or the sums that only depend on the time sampling ($CC_{\tau}$ and $SS_{\tau}$ or $CC$, $SS$, $CS$, and $D$, respectively). When adding the sine wave the mean changes
\begin{align*}
\overline{x} & =\sum w_{i}x_{i}=\sum w_{i}y_{i}+a\sum w_{i}\cos\omega_{0}t_{i}+b\sum w_{i}\sin\omega_{0}t_{i}\\
 & =\overline{y}+aC_{0}+bS_{0}.
\end{align*}
The sums $YC$ and $YS$ change as follows
\begin{align}
XC & =\sum w_{i}(x_{i}-\overline{x})\cos\omega t_{i}=YC+aCC_{0}+bCS_{0}\nonumber \\
XS & =\sum w_{i}(x_{i}-\overline{x})\sin\omega t_{i}=YS+aC_{0}S+bSS_{0}\nonumber \\
XX & =\sum w_{i}(x_{i}-\overline{x})^{2}\nonumber \\
   & =\sum w_{i}(y_{i}-\overline{y}+a(\cos\omega_{0}t_{i}-C_{0})+b(\sin\omega_{0}t_{i}-S_{0}))^{2}\nonumber \\
   & =YY+a^{2}C_{0}C_{0}+b^{2}S_{0}S_{0}+2aYC_{0}+2bY_{0}S_{0}+2abC_{0}S_{0}.\label{eq:XX}
\end{align}
As described in Sect.~\ref{sub:Upper-detection-limits}, it is more conservative to scan the power only at $\omega=\omega_{0}$ instead of the whole frequency range. This bypasses the calculation of sums comprising two different frequencies and is an enormous simplification. The sums of simulated data can be expressed by the sums for the original data. Therefore is not necessary to repeat the summation for the simulated data. When using the notation with $\tau$ this becomes $XC_{\tau}=YC_{\tau}+a_{\tau}CC_{\tau}$ and $XS_{\tau}=YS_{\tau}+a_{\tau}SS_{\tau}$ because $CS_{\tau}=0$.

The power response at the frequency where the sine wave was added is
\begin{align*}
p_{x}(\omega) & =\frac{1}{XX}\left[\frac{YC_{\tau}^{2}}{CC_{\tau}}+a^{2}CC_{\tau}+2aYC_{\tau}+\frac{YS_{\tau}^{2}}{SS_{\tau}}+b^{2}SS_{\tau}+2bYS_{\tau}\right]\\
 & =\frac{1}{XX}\left[p_{y}YY+XX-YY\right]=1-\frac{(1-p_{y})YY}{XX}.
\end{align*}

Now we are interested in the amplitude $A$ or variance $XX$ that is required for a given phase $\varphi$ to produce a desired power threshold $p_{x}(\omega)$, which corresponds to an FAP. The required variance is $XX=YY\frac{1-p_{y}}{1-p_{x}}$.

We obtain the required amplitude by solving the quadratic equation resulting from Eq.~(\ref{eq:XX}) ($a=A\cos\varphi$, $b=A\sin\varphi$)
\begin{align*}
0 & =a^{2}CC+b^{2}SS+2aYC+2bYS+2abCS-(XX-YY)\\
  & =A^{2}(CC\cos^{2}\varphi+SS\sin^{2}\varphi+2CS\cos\varphi\sin\varphi)\\
  & \quad+2A(YC\cos\varphi+YS\sin\varphi)-(XX-YY)\\
  & =\alpha A^{2}+2\beta A-\gamma\end{align*}
with the substitutions $\alpha=CC\cos^{2}\varphi+SS\sin^{2}\varphi+2CS\cos\varphi\sin\varphi$, $\beta=YC\cos\varphi+YS\sin\varphi$, and $\gamma=XX-YY=YY\frac{p_{x}-p_{y}}{p_{x}}$. This leads us to the amplitude
\[
A(\omega,\varphi)=-\frac{\beta}{\alpha}\underset{(-)}{+}\sqrt{\left(\frac{\beta}{\alpha}\right)^{2}+4\frac{\gamma}{\alpha}},
\]
which we consider as the amplitude detection limit for a fixed phase and for a given power threshold $p_{x}$ and which can be expressed by GLS sums for the original data. Probing a set of phases $\varphi$, we finally choose $\max_{\varphi}A(\omega,\varphi)$.

The second of the two solution is rejected, because we demand positive amplitudes ($A>0$, $\varphi\in[0,360\degr$). The terms $\alpha$ and $\gamma$ are always positive: $\gamma=YY\frac{p_{x}-p_{y}}{p_{x}}>0$ (as long as $p_{x}-p_{y}>0$) and $\alpha=CC\cos^{2}\varphi+SS\sin^{2}\varphi+2CS\cos\varphi\sin\varphi=CC_{\tau}\cos^{2}\varphi+SS_{\tau}\sin^{2}\varphi>0$.
\end{document}